\documentclass[11pt]{article}
\usepackage{latexsym}
\usepackage{amssymb}
\usepackage{amsmath}
\usepackage[hypertex]{hyperref}
\usepackage{graphicx}

\newcommand{\Qbar}{\overline{Q}}
\newcommand{\Mpl}{M_{\rm Pl}}
\newcommand{\ra}{\rightarrow}

\begin{document}

\pagestyle{empty}

\begin{flushright}
hep-ph/0501047 \\
\end{flushright}

\vspace{1.5cm}

\begin{center}
\mbox{\bf\LARGE Tripletless Unification in the Conformal Window}

\vspace*{1.5cm}
{\large Ryuichiro Kitano$^1$ and Graham D. Kribs$^{1,2}$} \\
\vspace*{0.5cm}

\mbox{$^1$\textit{School of Natural Sciences, Institute for Advanced Study, 
Princeton, NJ 08540}} \\[1mm]

\mbox{$^2$\textit{Department of Physics, University of Oregon, 
Eugene, OR 97403}}\footnote{\small On leave of absence.} \\

\vspace*{0.5cm}

\end{center}

\vspace*{1.0cm}

\begin{abstract}

A product SU(5) $\times$ Sp(4) grand unified model is proposed with 
no fundamental Higgs fields transforming under SU(5).  Higgs doublets
are instead embedded into a four dimensional representation of the 
Sp(4) gauge group, and hence there is no doublet-triplet splitting 
problem because there are no triplets.
The Sp(4) group contains enough matter to lie in the conformal window,
causing its gauge coupling to flow to a strongly-coupled infrared fixed-point 
at low energy, naturally preserving gauge coupling unification to 
percent level accuracy.  Yukawa couplings, including the top, 
arise through dimension five operators that are enhanced by the large 
anomalous dimension of the Higgs fields.
Proton decay mediated by dimension five operators is absent at the 
perturbative level.  It reappears, however, non-perturbatively due to 
Sp(4) instantons but the rate is suppressed by a high power of the ratio 
of the dynamical scale to the unification scale.
With gravity- or gaugino-mediated supersymmetry breaking, 
non-universal gaugino masses are predicted, 
satisfying specific one-loop renormalization group invariant relations.  
These predictions should be easily testable with the LHC and a 
linear collider.

\end{abstract} 

\newpage
\pagestyle{plain}

\section{Introduction}

The intersection of the three Standard Model (SM) gauge couplings 
at high energy with weak scale supersymmetry has long remained a 
tantalizing hint for a supersymmetric grand unified theory (GUT).
Grand unification simplifies the matter content, by embedding matter
into much simpler GUT representations, and simplifies the gauge group 
structure, by embedding the standard model into a simple GUT group. 
Minimal non-supersymmetric and supersymmetric simple group unification 
does, however, face a severe fine-tuning problem associated with the 
Higgs sector.  Embedding Higgs doublets into GUT representations requires 
color triplet partners.  The color triplet partners must be heavy, 
to prevent Higgs-mediated proton decay and preserve gauge coupling
unification, while the the Higgs doublets must be light to obtain
electroweak breaking at the weak scale.  This is the infamous
doublet-triplet splitting problem.

In supersymmetry, the doublet-triplet splitting problem is even more 
enigmatic because the fermionic partners to the color triplet Higgs fields 
lead to proton decay through dimension five operators at a rate somewhat 
faster than the present experimental bound.  In minimal SU(5), for example,
the mass scale of the color triplet Higgsinos must be several times larger 
than the GUT scale \cite{Goto:1998qg, Murayama:2001ur}.
There are several approaches to address this problem and a vast
accompanying literature.  (See for example in four dimensions 
\cite{Dimopoulos:1981xm,Masiero:1982fe,Antoniadis:1987dx,Babu:1993we,Inoue:1985cw,Yanagida:1994vq}, 
extra dimensions \cite{Kawamura:2000ev,Hebecker:2001wq}, 
deconstruction \cite{Csaki:2001qm,Witten:2001bf}, 
and string theory \cite{Candelas:1985en}.)

Our proposal to solve doublet-triplet splitting is to dispose
entirely of Higgs triplets by simply removing the complete GUT Higgs 
representation out of the theory.  Of course Higgs doublet fields must 
somehow reappear in the model so that electroweak symmetry can be broken.  
The mechanism we employ to accomplish this is to embed the Standard Model 
into a diagonal subgroup of a GUT group and another simple group.  
Something similar was first 
footnoted by Yanagida \cite{Yanagida:1994vq,Hotta:1996pn} who proposed 
SU(5) $\times$ SU(2)$_H$ $\times$ U(1)$_H$.  (Later the proton decay 
signal of this model was examined in Ref.~\cite{Ibe:2003ys}.)  
The idea is that the would-be Higgs fields begin their life as 
${\bf 2} + {\bf 2}$ of SU(2)$_H$, without the need of any color partners, 
and become ordinary Higgs doublets only after the product group diagonally 
breaks to the SM\@.  This is what we call a ``tripletless'' GUT model. 

This is similar in some respects to the missing partner mechanism
where doublet-triplet splitting is accomplished by introducing additional 
partners whose sole purpose is to marry off the color triplet Higgs fields.
Dimension five operators for proton decay are suppressed or 
forbidden by enforcing appropriate auxiliary symmetries, such as 
global U(1)$_{PQ}$, that restrict the form of the allowed mass terms.
Missing partner models are most easily constructed with product (GUT) 
gauge theories.  The simplest model is based on SU(5) $\times$ SU(3)$_H$ 
$\times$ U(1)$_H$ proposed by Yanagida \cite{Yanagida:1994vq} and studied 
in detail in Refs.~\cite{Hotta:1995cd,Hisano:1995hc,Hotta:1996pn,Izawa:1997he,Imamura:2001es}.  
The partners of the color triplet Higgs fields are 
a ${\bf 3} + {\bf \bar{3}}$ of SU(3)$_H$ in the model.  After the GUT 
breaks, the color triplet Higgs fields obtain mass by pairing off with 
these partners while the doublets do not acquire mass due to the lack of 
suitable partners.  Missing partner models that marry off color triplets 
with other fields in the process of product group diagonal breaking we 
call ``tripletfull'' GUT models.  

Tripletfull and tripletless product group models have the virtue 
that they completely solve the doublet-triplet splitting problem and can 
easily solve the triplet-induced dimension-5 proton decay problem.  
But they too face problems of their own creation.  First, the gauge 
couplings do not generically unify.  Instead, approximate unification 
emerges only if the hypergroups' couplings are large, greater than of 
order a few to obtain the intersection of the gauge couplings at the 
GUT scale to within a few percent.  Large values of these couplings, 
however, is itself problematic.  A strongly-coupled U(1)$_H$ is a disaster, 
since the U(1) is not asymptotically free.  The U(1) hits a Landau pole only
slightly above the GUT scale, causing the effective theory to break down.
Furthermore, the other hypercolor groups may also not be asymptotically free 
(the SU(2)$_H$ coupling in the above tripletless model is an example).
Finally, the U(1)$_H$ factor also spoils the automatic prediction of 
charge quantization.

All the above problems can be avoided in models with the SU(5) $\times$
$G$ gauge interaction with $G$ being a simple, asymptotically-free group. 
Asymptotic freedom allows a large value of the gauge coupling at the 
GUT scale without any UV difficulties.  This general approach, however, 
may have a coincidence problem.  There is no dynamical explanation why the 
$G$ group gauge coupling is large at the same scale as the GUT breaking 
scale (determined by other parameters in the superpotential), and this 
means a modest fine-tuning appears to be unavoidable.

In this paper we propose the first tripletless GUT model with all
high energy groups asymptotically free based on SU(5) $\times$ Sp(4).
We show that the Sp(4) gauge interaction has a strongly-coupled infrared 
fixed point that automatically solves the coincidence problem.  
The IR fixed point behavior also enhances the dimension-5 Yukawa
couplings, allowing us to obtain the top Yukawa from a higher
dimensional operator with an ${\cal O}({\rm few})$ coupling.
Proton decay through dimension five operators is perturbatively
forbidden, but reappears non-perturbatively at a suppressed level due
to quantum effects of Sp(4).  Given supersymmetry breaking communicated 
through  gravity or gaugino mediation, non-universal gaugino masses are 
predicted, satisfying specific one-loop renormalization group invariant 
relations.  These predictions should be easily testable with the LHC and 
a linear collider.

Finally, we remark that there is a tripletfull GUT model in this class 
of theories, the SO(10) $\times$ SO(6)$_H$ model \cite{Hotta:1996qb}.  
This theory, with 11 flavors transforming under SO(6)$_H$, is within but 
right on the edge of the conformal window.  If the SO(6) coupling is 
already at its IR fixed point at the Planck scale, there is no coincidence 
problem.  However, this model does require numerous auxiliary fields to 
achieve SO(10) breaking.
As we will see below, our tripletless model is much simpler in field
content in part because higher dimensional operators receive a significant 
enhancement through the large anomalous dimensions of the fields.

\section{The SU(5) $\times$ Sp(4) Model}

The Higgs sector of our model consists of an Sp(4) gauge theory with 
six flavors, $T_1$ and $T_2$, $Q^i$ and $\Qbar_i$ where $i=1\ldots5$.  
SU(5) is a gauged subgroup of the global SU(12) flavor symmetry.
The particle content of the model is summarized in Table~\ref{tab:content}.
\begin{table}
\begin{center}
\begin{tabular}{c|cc}
 & SU(5) & Sp(4)  \\ \hline
 $Q_\alpha^i$         & {\bf 5} & {\bf 4} \\ 
 ${\Qbar}_{\alpha,i}$ & ${\bf \overline{5}}$ & {\bf 4}  \\ 
 $T_{1, \alpha}$      & {\bf 1} & {\bf 4}  \\ 
 $T_{2, \alpha}$      & {\bf 1} & {\bf 4}  \\
\end{tabular}
\end{center}
\caption{The particle content of our SU(5) $\times$ Sp(4) tripletless 
product GUT model.  Matter (not shown) is embedded into the usual 
$({\bf 10},{\bf 1}) \oplus ({\bf \overline{5}},{\bf 1})$ 
representations of SU(5).}
\label{tab:content}
\end{table}
The tree level superpotential is given by
\begin{eqnarray}
 W &=& m (Q \cdot \Qbar) 
       - \frac{1}{M_1} (Q \cdot \Qbar) (Q \cdot \Qbar)
       - \frac{1}{M_2} (Q^i \cdot \Qbar_j) (Q^j \cdot \Qbar_i)
       - \frac{1}{M_3} (Q^i \cdot Q^j) (\Qbar_j \cdot \Qbar_i)
       + \cdots \nonumber \\
&&{}   + \frac{1}{M_T} (Q^i \cdot T_2) (\Qbar_i \cdot T_2) \ ,
\label{eq:superpotential}
\end{eqnarray}
where $i,j,\ldots$ represent SU(5) indices.  We use the notation 
$(A^i \cdot B_j)$ to denote the Sp(4) contraction 
$A^i_\alpha B_{\beta,j} J^{\alpha\beta}$, where $\alpha, \beta$ 
represent the Sp(4) indices.  
The matrix $J^{\alpha \beta} = i \sigma_2 \otimes {\bf 1}_2$ is the 
invariant antisymmetric tensor of Sp(4).
The contraction $(A \cdot B)$ without 
explicit SU(5) indices corresponds to $(A^i \cdot B_j) \delta^j_i$.
We have assumed that $T_1$ does not appear in the superpotential, 
since it will become a pair of Higgs doublets after the GUT breaks.
The extra field $T_2$ is necessary to avoid the Witten anomaly
\cite{Witten:1982fp}.
We omit all other possible operators as they will not affect 
our subsequent analysis.

Superpotential terms such as $(Q \cdot T_1) (\Qbar \cdot T_1)$ (and
mixed terms involving $T_1$ and $T_2$) must not be written with order
one coefficients because they induce a GUT scale $\mu$-term for the
Higgs fields.  Symmetries can easily justify the lack of such terms.
For example, impose a global U(1) symmetry under which $T_1$ has charge
unity and neither of $Q$, $\Qbar$, nor $T_2$ are charged.  Yukawa
couplings are allowed with the charge assignment of ${\bf 10}: -1/2$ and
${\bf \bar{5}}: -1/2$.  This U(1) symmetry is, however, anomalous and
broken by Sp(4) instantons. Nevertheless it is straightforward to modify
the model by adding spectators charged under Sp(4) and U(1) so as to
cancel the mixed anomaly.  For example, adding $N$ flavors of Sp(4) with
a U(1) charge $-1/(2N)$ cancels the Sp(4)-Sp(4)-U(1) anomaly.  The extra
flavors will obtain mass if the U(1) symmetry is broken at high energy
by vacuum expectation values (vevs) of fields $\phi$ and $\bar{\phi}$
with charge $1/N$ and $-1/N$.  The lowest-dimensional gauge-invariant
operator leading to a $\mu$-term is $\phi^{2 N} (Q \cdot T_1) (\Qbar
\cdot T_1)/M^{2 N - 1}$, and can be suppressed to the weak scale or
below by taking $N$ large enough and/or the ratio $\langle \phi
\rangle/M$ small enough.

We now analyze the moduli space of the theory.  At the origin where
$Q_\alpha^i = \Qbar_{\alpha, i} = 0$, the low energy effective theory is
an Sp(4) gauge theory with one flavor that is known to generate a
non-perturbative superpotential with a runaway direction and therefore
no ground state.  Hence, $Q$ and $\Qbar$ must acquire vevs so that the
GUT group breaks down~\cite{Hotta:1996qb}.
Interestingly, as we see later, $Q$ and $\Qbar$ also cannot acquire an
expectation value in only one of its components as this would leave an 
Sp(2) theory with a dynamically generated run-away superpotential.
Hence, $Q$ and $\Qbar$ must acquire vevs for two components, uniquely
selecting the vacuum with rank$\langle Q \rangle=2$ that suggests the
SM is the preserved diagonal subgroup gauge
symmetry.\footnote{Note that the SM is uniquely chosen if the
superpotential is minimal, i.e., Eq.~(\ref{eq:superpotential}) without
the `$\cdots$' terms.}

First we analyze the theory classically, showing how doublet-triplet 
splitting is achieved.  Later we discuss in detail the non-perturbative
corrections.  The potential is minimized with
\begin{eqnarray}
 Q = \left(
\begin{array}{llll}
 0 & 0 & 0 & 0 \\
 0 & 0 & 0 & 0 \\
 0 & 0 & 0 & 0 \\
 v & 0 & 0 & 0 \\
 0 & v & 0 & 0 
\end{array}
\right)\ ,\ \ \ 
 \Qbar = \left(
\begin{array}{llll}
 0 & 0 & 0 & 0 \\
 0 & 0 & 0 & 0 \\
 0 & 0 & 0 & 0 \\
 0 & 0 & v & 0 \\
 0 & 0 & 0 & v 
\end{array}
\right)\ ,\ \ \
T_1 = T_2 = 0 \ ,
\label{eq:vacuum}
\end{eqnarray}
where
\begin{eqnarray}
 v = \sqrt{\frac{m M_1 M_2}{2 M_1 + 4 M_2}} \; .
\end{eqnarray}
The vanishing components of the matrices are ensured by the Sp(4) and
SU(5) $D$-flat conditions.  
In the vacuum given by Eq.~(\ref{eq:vacuum}), SU(5) $\times$ Sp(4) is 
broken down to the standard model gauge group.  
The SU(2)$_L$ $\times$ U(1)$_Y$ electroweak group is the diagonal 
subgroup of SU(2) $\times$ U(1) in SU(5) and that in Sp(4).  
All of the pseudo-Goldstone components in $Q$, $\Qbar$, and $T_2$ 
acquire mass (as we show in detail below) while $T_1$ remains massless.  
$T_1$ becomes the Higgs doublet fields of the minimal supersymmetric 
standard model (MSSM).  The doublet-triplet splitting problem is 
trivially solved as there are no colored Higgs particles.

Matching the SM gauge couplings $g_{1,2,3}$ to the high energy couplings 
$g_5$ [SU(5)] and $g_4$ [Sp(4)] we obtain the tree-level relations
\begin{equation}
\frac{1}{g_3^2} = \frac{1}{g_5^2} \quad , \quad
\frac{1}{g_2^2} = \frac{1}{g_5^2} + \frac{2}{g_4^2} \quad , \quad
\frac{1}{g_1^2} = \frac{1}{g_5^2} + \frac{6}{5 g_4^2} \; .
\label{matching}
\end{equation}
The numerical factors are determined by the embedding of SU(2) $\times$ U(1) 
in Sp(4).  Here $g_1$ is in the usual GUT normalization where 
$g_Y^2 = 3/5 g_1^2$. 
Obtaining gauge coupling unification at ${\cal O}(1\%)$ level requires 
the gauge coupling of Sp(4) to be $g_4 \gtrsim 7$ at the GUT scale. 

Such a large coupling for Sp(4) implies that we must consider the 
non-perturbative effects of this strongly-coupled gauge interaction. 
The most convenient way to analyze the theory non-perturbatively is
to first rewrite the superpotential Eq.~(\ref{eq:superpotential}) 
in terms of renormalizable interactions:
\begin{eqnarray}
 W &=& m (Q^I \cdot \Qbar_I)
+ \lambda_{24} X^j_i (Q^i \cdot \Qbar_j) 
+ \lambda_S S (Q \cdot \Qbar)
+ m_{24} X^i_j X^j_i
+ m_S S^2
+ \tilde{m}^2 (S - \sqrt{15} \, {\rm tr} \, t^{24} X)
\nonumber \\
&&{} 
+ k \tilde{H}^i (\Qbar_i \cdot T_2 )
+ \bar{k} \overline{\tilde{H}}_i (Q^i \cdot T_2)
+ m_5 \tilde{H}^i \overline{\tilde{H}}_i \; .
\label{renormalizable-superpotential}
\end{eqnarray}
Here $I,J=1,2,3$, and $S$, $X$, $\tilde{H}$, and $\overline{\tilde{H}}$
are massive SU(5) singlet, adjoint, fundamental and anti-fundamental
fields, respectively, that do not acquire vevs.
This superpotential is equivalent to Eq.~(\ref{eq:superpotential}) upon
integrating out the heavy fields.  Nevertheless, this superpotential
shows that our model is simply a mass deformed Sp(4) gauge theory with
six flavors.  Integrating out the three massive flavors gives a low
energy effective theory described by $Q^A_\alpha$, $\Qbar_{A,\alpha}$,
$T_1$ and $T_2$ with $A=4,5$ and spectators $X$, $S$, $\tilde{H}$ and
$\overline{\tilde{H}}$ which can be thought of as mass deformations,
i.e., the mass deformed three flavor theory.
The non-perturbative effect of this theory is well-known to be a theory
with quantum modified moduli space.
By symmetries, holomorphy and various limits, we cannot write down a
superpotential.  The flavor symmetry suggests that the non-perturbative
superpotential appears with a combination of ${\rm Pf}\, M$, where $M$
is the meson fields made of two of $Q$, $\Qbar$, and $T$'s, but it is
forbidden by the non-anomalous $R$-symmetry as
$R(M)=0$~\cite{Seiberg:1993vc}.
Since $T_1$ could only appear in the superpotential by the
non-perturbative effect, mass terms for the Higgs-doublet fields are not
generated even though there is no unbroken symmetry which protects the
masslessness.  In this low energy effective theory, the moduli space is
modified by quantum effects.  In particular, the classical constraint
${\rm Pf} \, M = 0$ is modified to ${\rm Pf} \, M = \Lambda^6$, where
$\Lambda$ is the dynamical scale of the three flavor Sp(4) theory.  The
constraint forces an SU(5)-singlet composite direction $M_{TT} \sim (T_1
\cdot T_2)$ to obtain a vev of the order of
\begin{eqnarray}
 \langle M_{TT} \rangle \sim \frac{\Lambda^6}{v^4}\ .
\label{eq:t1t2vev}
\end{eqnarray}
The other directions such as $(\bar{Q} \cdot T_1 )(Q \cdot T_2)$ are
protected against acquiring vevs by the superpotential terms with 
$\tilde{H}$ and $\overline{\tilde{H}}$.
As we will see, this non-perturbative effect in Eq.~(\ref{eq:t1t2vev}) is
important in our discussion of proton decay.  Finally, notice that the
vacuum with rank $\langle Q \rangle = 2$ is uniquely chosen dynamically.
A similar analysis with above shows that, in the vacuum with
rank$\langle Q \rangle = 1 (0)$, the effective theory is an Sp(4) gauge
theory with two (one) flavors.  Such small $N_f$ theories are known to
dynamically generate a run-away superpotential for the massless flavors
causing the vacuum to be unstable.

The most important non-perturbative quantum effect of our strongly-coupled 
Sp(4) is the existence of an infrared fixed point.  
Above the scale $m$, the theory is an Sp($2 N_c$) gauge theory 
with $2 N_c = 4$ and $N_f = 6$, which is
known to be in the conformal window $\frac{3}{2} (N_c + 1) < N_f < 3
(N_c + 1)$ with a non-trivial strongly-coupled fixed-point of the 
renormalization group \cite{Intriligator:1995ne}.  
There are several important consequences.
First, the gauge coupling is naturally large, ${\cal O}(4 \pi)$, 
which ensures the gauge couplings unify to excellent accuracy.
Next, there is no coincidence between the scale where the gauge
coupling becomes strong and the GUT breaking scale.  In particular,
they are both determined the supersymmetric mass, $m$, that effectively
sets the dynamical scale $\Lambda \sim m$.  
Finally, the coupling constants in the superpotential obtain a large
enhancement from the large anomalous dimensions of the Sp(4) fields.
The anomalous dimension of the meson operator is 
\begin{equation}
\gamma^* = 1 - 3 \frac{N_c + 1}{N_f} = -\frac{1}{2}
\end{equation}
which enhances the coefficients of the operators at low energy such as
\begin{eqnarray}
 m (\mu) = m (\mu_0) \left(
\frac{\mu}{\mu_0}
\right)^{-\frac{1}{2}} \ ,\ \ \ 
\frac{1}{M_X (\mu)} =
\frac{1}{M_X (\mu_0)} \left(
\frac{\mu}{\mu_0}
\right)^{-1} \ .
\label{MXenhance}
\end{eqnarray}
where $M_X$ stands for $M_1$, $M_2$, $M_3$, and $M_T$ in
Eq.~(\ref{eq:superpotential}), and $\mu_0$ is the scale where the gauge
coupling constant reaches ${\cal O}(4 \pi)$.
Clearly, the non-renormalizable operators in
Eq.~(\ref{eq:superpotential}) are not necessarily suppressed by the
Planck scale but could be only suppressed by the GUT scale, if the 
Sp(4) group is already strongly coupled at the Planck scale.

The Yukawa interactions between matter and Higgs fields arise from higher
dimensional operators 
\begin{eqnarray}
 W_{\rm Yukawa} = 
  \frac{1}{M_{U}} \epsilon_{ijklm} {\bf 10}^{ij} {\bf 10}^{kl} (Q^m \cdot T_1)
+ \frac{1}{M_{D}} {\bf 10}^{ij} {\bf \bar{5}}_i (\Qbar_j \cdot T_1) \ ,
\end{eqnarray}
where ${\bf 10}$ and ${\bf \bar{5}}$ are the matter fields and the
generation indices have been suppressed.
For the top quark, the Yukawa coupling evaluated at the GUT scale
is roughly $\lambda_t \sim 0.5 - 0.7$ (depending on $\tan\beta$)
which appears to require an unnaturally small $M_{U}$.
However, the large anomalous dimensions of the Sp(4) fields causes
an enhancement of up to ${\cal O}[(\Mpl / M_{\rm GUT})^{1/2}] \sim 10$.  
We need only a slightly large coefficient of 
the Planck suppressed operator, of order 5, to obtain the observed 
top Yukawa coupling at the weak scale.  It is also possible to enhance 
the top Yukawa coupling by introducing a pair of massive fields with 
${\bf 5} + {\bf \bar{5}}$ of SU(5) with renormalizable interactions
that, after integrating out these fields, generates a top Yukawa.
We will not pursue this (admittedly ad hoc) extension of our model, 
but note nevertheless that it does not invalidate our analysis of the 
masslessness of the Higgs doublet fields.

The enhancement of the mass parameter suggests that the original mass
scale of the model is lower than the GUT scale.
If the neutrino masses are explained by the seesaw mechanism, the
natural scale for the right-handed neutrino masses are the lower
original scale because there is no enhancement. It provides a natural
explanation for the smallness of the right-handed neutrino masses which
is necessary to reproduce the observed neutrino masses.

Since there are no colored Higgs fields, dimension five operators leading
to proton decay are absent at the classical level.  However, 
non-perturbative effects can reintroduce these operators with 
power-suppressed coefficients.  After integrating out the heavy fields, 
we obtain operators such as
\begin{eqnarray}
 \frac{ \lambda_u \lambda_d }{ M_T m^2 v^2 } \epsilon_{ijklm}
{\bf 10}^{ij} {\bf 10}^{kl} {\bf 10}^{mn} {\bf \bar{5}}_n
(T_1 \cdot T_2) (T_1 \cdot T_2) \ ,
\end{eqnarray}
where $\lambda_u$ and $\lambda_d$ are the Yukawa coupling constants.  
Since the combination of $(T_1 \cdot T_2)$ acquires a vev from the
quantum modified constraint, Eq.~(\ref{eq:t1t2vev}), we obtain
\begin{eqnarray}
 W_{\rm eff} =  \frac{ f_u f_d \Lambda^{12} }{ M_T m^2 v^{10} } \ 
\epsilon_{ijklm} {\bf 10}^{ij} {\bf 10}^{kl} {\bf 10}^{mn} {\bf \bar{5}}_n \ .
\end{eqnarray}
The constraint from the proton lifetime requires the effective colored
Higgs mass of $(M_T m^2 v^{10}) / \Lambda^{12}$ to be larger than about
$10^{17}$~GeV (see, e.g., Ref.~\cite{Murayama:2001ur}). 
This is easily satisfied by taking the dynamical scale $\Lambda$ to be 
slightly smaller than the GUT scale $v$.  This is naturally realized in 
our model since the GUT scale is ${\cal O}(\sqrt{m M})$ while 
$\Lambda \sim m$.

\section{Threshold Corrections to Gauge Coupling Unification}

Here we discuss the threshold corrections to gauge coupling unification.
The matter content of our model, decomposed into fields transforming 
under the SM group,
\begin{eqnarray}
Q &=& \left[ \begin{array}{cc} 
          ({\bf 3},{\bf 2})_{-5/6} & ({\bf 3},{\bf 2})_{1/6} \\
          ({\bf 1},{\bf 3})_0 \oplus ({\bf 1},{\bf 1})_0 & 
          ({\bf 1},{\bf 3})_1 \oplus ({\bf 1},{\bf 1})_1 \end{array} \right] \\
\Qbar &=& \left[ \begin{array}{cc} 
          ({\bf \bar{3}},{\bf 2})_{-1/6} & ({\bf \bar{3}},{\bf 2})_{5/6} \\
          ({\bf 1},{\bf 3})_{-1} \oplus ({\bf 1},{\bf 1})_{-1} & 
          ({\bf 1},{\bf 3})_0 \oplus ({\bf 1},{\bf 1})_0 \end{array} \right] \\
T_1 &=& \left[ \; ({\bf 1},{\bf 2})_{-1/2} \;\; ({\bf 1},{\bf 2})_{1/2} \; 
        \right] \\
T_2 &=& \left[ \; ({\bf 1},{\bf 2})_{-1/2} \;\; ({\bf 1},{\bf 2})_{1/2} \; 
        \right]
\end{eqnarray}
where our notation for the fields is to write their quantum numbers as 
$[{\rm SU(3)}_c, \, {\rm SU(2)}_L]_{{\rm U(1)}_Y}$.
Upon SU(5) $\times$ Sp(4) breaking to SU(3)$_c$ $\times$ SU(2)$_L$ $\times$
U(1)$_Y$, there are $24 + 10 - 12 = 22$
broken generators.  These massive gauge bosons include
the $X,Y$'s of SU(5) that eat the $({\bf 3},{\bf 2})_{-5/6}$ and
$({\bf \bar{3}},{\bf 2})_{5/6}$ fields;
the broken Sp(4) generators that eat the $({\bf 1},{\bf 3})_{\pm 1}$;
and the broken linear combinations of the remaining Sp(4) generators and 
the SU(2) $\times$ U(1) subgroup of SU(5) that absorb one 
$({\bf 1},{\bf 3})_0$ and one $({\bf 1},{\bf 1})_0$ fields.  
$T_1$ becomes the a pair of SM Higgs doublets that survive to the weak scale.  
This leaves several pseudo-Goldstone fields that acquire mass ${\cal O}(m)$ 
only from the higher dimensional operators in Eq.~(\ref{eq:superpotential}).
Specifically, the uneaten $({\bf 1},{\bf 3})_0$ field  
acquires a mass of order $v^2/M_1 + v^2/M_2$;
the $({\bf 3},{\bf 2})_{1/6}$ and $({\bf \bar{3}},{\bf 2})_{-1/6}$
fields pair up with a mass of order $v^2/M_2 + v^2/M_3$; 
finally the $({\bf 1},{\bf 1})_{\pm 1}$ fields pair up with a mass
of order $v^2/M_3$.  Finally, the second set of Higgs doublets 
from $T_2$ pair up with a mass of order $v^2/M_T$.

\begin{figure}[t]
\begin{center}
\setlength{\unitlength}{1mm}
\begin{picture}(120,60)
\put(20,0){\vector(0,1){60}}

\put( 1, 9){$\sim v^2/\Mpl$}
\put(19,10){\line(1,0){2}}
\put(25,10){\line(1,0){30}}
\put(25,12){pseudo-Goldstones}

\put( 1,29){$g_5 v \sim g_4 v$}
\put(19,30){\line(1,0){2}}
\put(25,30){\line(1,0){30}}
\put(25,32){$X,Y$; broken Sp(4)}

\put(11,49){$\Mpl$}
\put(19,50){\line(1,0){2}}

\put(80, 0){\vector(0,1){60}}

\put(85,29){\line(1,0){30}}
\put(85,26){pseudo-Goldstones}

\put(68,29){$\sim g_5 v$}
\put(79,30){\line(1,0){2}}

\put(85,30){\line(1,0){30}}
\put(95,30.5){$X,Y$}

\put(68,33){$\sim g_4 v$}
\put(79,34){\line(1,0){2}}

\put(85,34){\line(1,0){30}}
\put(88,36){broken Sp(4)}

\put(71, 49){$\Mpl$}
\put(79,50){\line(1,0){2}}

\end{picture}
\end{center}
\caption{A sketch of the mass spectrum is shown for a weakly-coupled 
SU(5) $\times$ Sp(4) theory (left side), and for our model with a 
strongly-coupled Sp(4) that is conformal up to the Planck scale with 
the maximal anomalous dimension enhancement (right side).}
\label{spectrum-fig}
\end{figure}
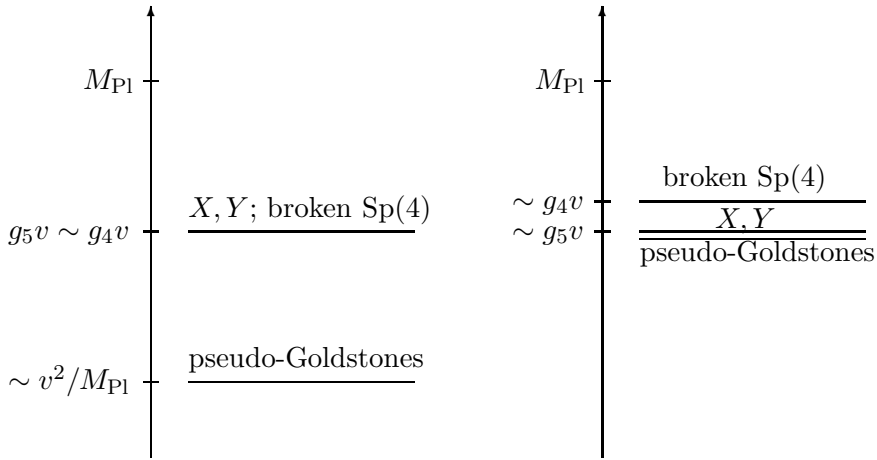

Above the scale of the pseudo-Goldstone fields, but below the GUT breaking
scale, the one-loop beta function coefficients of the SM couplings are 
shifted by an amount
\begin{equation}
\Delta b_1 = 2 \quad , \quad \Delta b_2 = 6 \quad , \quad
\Delta b_3 = 2 \; .
\end{equation}
In a weakly-coupled product gauge theory, taking the pseudo-Goldstone fields
to have roughly the same mass scale $v^2/M \sim 10^{-2} v$, these shifts
in the beta functions induce a very significant threshold correction
to gauge coupling unification.\footnote{The higher dimensional
operators could have different coefficients and therefore the pseudo-Goldstone
fields could have different masses.  In particular, the
$({\bf 1},{\bf 1})_{\pm 1}$ fields could be easily made much
lighter than the other pseudo-Goldstone fields, mitigating some 
of effects of this threshold correction.}
This would-be disaster is averted due to the large anomalous dimensions 
of the strongly-coupled Sp(4) that provide a large enhancement to the 
mass parameters of the model, as we found in Eq.~(\ref{MXenhance}).  
Let us consider the maximum enhancement, when the Sp(4) gauge interaction 
is conformal up to the Planck scale.  Starting with dimension-5 operators 
that are $1/\Mpl$ suppressed, the conformal enhancement causes 
$1/\Mpl (M_{\rm GUT}/\Mpl)^{-1} \ra 1/M_{\rm GUT}$, and therefore 
the masses of the pseudo-Goldstone fields are increased up to roughly 
the GUT scale.  Obviously no large threshold correction from these 
fields is expected.  This required assuming the Sp(4) group was 
conformal up to the Planck scale, which is consistent with maximizing 
the enhancement of the top Yukawa.
We sketch the spectrum of this theory in a purely weakly-coupled
case and in the strongly-coupled, conformal up to the Planck scale
case in Fig.~\ref{spectrum-fig}.

With Sp(4) in the conformal window, the correction to SM gauge coupling
from tree-level matching, Eqs.~(\ref{matching}), is also negligible.
This assumes of course that the infrared fixed-point value $g_4 \gtrsim 7$.
This cannot be verified perturbatively, but as an aside we remark
that the two-loop correction to the Sp(4) gauge coupling running
is the same as the one-loop correction when $g_4 = 2 \pi$, 
which roughly satisfies our constraint.
A large Sp(4) coupling also causes the massive gauge bosons to be
split in mass, namely $m_{X,Y} \sim g_5 v \equiv M_{\rm GUT}$ while
the the broken Sp(4) generators have a mass of order $g_4 v$.
The effect of this threshold correction is equivalent to a (fake)
redefinition of the coupling $g_4$ in the tree-level matching,
and so causes no new issues.

\section{Experimental Signals}

The dominance of proton decay through dimension six operators is
the smoking gun of tripletless models, as well as most proposals
to solve the doublet-triplet splitting problem in supersymmetry.
In our model, the GUT scale (scale of $X,Y$ gauge bosons) is at
or slightly higher than the scale in minimal SU(5), which means the 
proton decay rate through $X,Y$ exchange is expected to be even 
(slightly) longer than would be predicted in minimal SU(5) GUT\@.
Proton decay through dimension-five operators is possible but
extremely unlikely due to the severe sensitivity to the ratio
$\Lambda/v$.  This does not bode well for testing our model through
proton decay experiments.

Supersymmetric product group models do have one very interesting signal:  
gaugino mass non-universality \cite{Arkani-Hamed:1996jq}.
Suppose that the mechanism of supersymmetry breaking is 
UV sensitive, leading to gaugino masses for both SU(5) gauginos
and Sp(4) gauginos through operators such as
\begin{equation}
\int d^2\theta \, \left[ f_5(S) W_\alpha^5 W^{5,\alpha} 
+ f_4(S) W_\alpha^4 W^{4,\alpha} \right] \; ,
\end{equation}
where $f_5(S)$ and $f_4(S)$ are gauge kinetic functions that acquire 
expectation value for both the scalar and auxiliary components.
These operators appear in supergravity mediation as well as gaugino
mediation \cite{Kaplan:1999ac}.
The scalar components of the gauge kinetic functions give the gauge coupling
constants $[f_5]_S = 1/(2g_5^2)$ and $[f_4]_S = 1/(2g_4^2)$, whereas
the gaugino masses are given by $[f_5]_F = M_5/g_5^2$ and 
$[f_4]_F = M_4/g_4^2$.
This leads to tree-level matching relations for the
SM gauginos
\begin{eqnarray}
\frac{M_3}{\alpha_3} &=& \frac{M_5}{\alpha_5} \\
\frac{M_2}{\alpha_2} &=& \frac{M_5}{\alpha_5} + 2 \frac{M_4}{\alpha_4} \\
\frac{M_1}{\alpha_1} &=& \frac{M_5}{\alpha_5} 
+ \frac{6}{5} \frac{M_4}{\alpha_4} \; .
\end{eqnarray}
The deviations from gaugino mass universality are proportional to 
$M_4/\alpha_4$.
If the auxiliary components of the gauge kinetic functions $f_5$ and $f_4$ 
are comparable, the deviation from universality is expected to be large,
although this is UV-dependent.

Notice that the three SM gaugino masses depend on only two high energy 
parameters.  This implies the prediction 
\begin{equation}
\frac{M_1}{\alpha_1} - \frac{3}{5} \frac{M_2}{\alpha_2} 
                     - \frac{2}{5} \frac{M_3}{\alpha_3} = 0 \; ,
\end{equation}
that is renormalization group invariant to one-loop, and therefore
valid at the weak scale to within a few percent accuracy \cite{Kribs:1998rb}.
This appears to be a unique prediction of our SU(5) $\times$ Sp(4) 
product GUT model.  Certainly there is no such prediction for the 
product GUT models involving SU(5) $\times$ SU(3)$_H$ $\times$ U(1)$_H$ 
since the U(1) normalization is unknown.  However, our relation happens 
to be precisely the same as predicted by the SO(10) $\times$ SO(6) 
tripletfull model \cite{Kurosawa:1999bh}\footnote{We thank Y. Nomura 
for pointing this out to us.}.  This happens despite, for example, 
SU(2)$_L$ coming from a diagonal subgroup of SU(5) $\times$ Sp(4) 
in our model, whereas it comes from purely SO(10) in the model of 
Ref.~\cite{Hotta:1996qb}.
This means that to unambiguously test our model against other product
group models we must examine the individual ratios of gaugino masses.  
With our model, we find the ratios
\begin{eqnarray}
1 &<& \left. \frac{M_2/\alpha_2}{M_1/\alpha_1} 
    \right|_{{\rm SU(5)} \times {\rm Sp(4)}} \; < \; \frac{5}{3} \\
1 &>& \left. \frac{M_3/\alpha_3}{M_2/\alpha_2} 
    \right|_{{\rm SU(5)} \times {\rm Sp(4)}} \; > \; 0
\end{eqnarray}
where the left(right)-hand side is obtained in the limit $M_4 \ra 0$ 
($M_5 \ra 0$).  In models in which SU(2) does \emph{not} arise as a diagonal
subgroup, e.g.\ tripletfull models such as the SO(10) $\times$ SO(6) model,
these ratios always lies in the range
\begin{eqnarray}
1 &>& \left. \frac{M_2/\alpha_2}{M_1/\alpha_1} \right|_{\rm tripletfull} 
  \; > \; 0 \\
1 &<& \left. \frac{M_3/\alpha_3}{M_2/\alpha_2} \right|_{\rm tripletfull} 
  \; < \; \infty
\end{eqnarray}
where again the left(right)-hand side is obtained in the limit the
hypercolor gaugino mass(es) go to zero (the GUT gaugino mass goes to zero).
The Bino-to-Wino mass ratio may be the best bet experimentally, since it is 
expected to be tested to high accuracy at a linear collider by looking 
at the endpoint distribution in chargino decay \cite{Aguilar-Saavedra:2001rg}.
Mass ratios involving the gluino inevitably involve larger QCD 
uncertainties, but may still yield important information.

\section{AdS interpretation?}

Our model was constructed purely as a four-dimensional strongly-coupled
supersymmetric product group theory.  Nevertheless, it is intriguing to
speculate about a possible dual five-dimensional AdS interpretation to 
our CFT theory.  We emphasize that there is by no means a certainty that 
such an AdS interpretation exists, but it is interesting to sketch the 
outcome of reversing the AdS/CFT dictionary.  Our analysis is in the
spirit of \cite{Arkani-Hamed:2000ds}, and perhaps most closely linked 
with the CFT interpretation \cite{Nomura:2003du}
(see also \cite{Contino:2003ve})
of warped Higgsless theories \cite{Csaki:2003zu}.

The region between Planck and the GUT scales translates into a slice of 
AdS space in which the endpoints are interpreted as Planck and GUT branes.  
Our CFT has a large global symmetry, part of which is gauged.  
The weakly gauged SU(5) subgroup of the SU(12) global symmetry translates
into the bulk of AdS containing the SU(5) gauge symmetry.  
Matter fields in our CFT are elementary fields, corresponding to Planck
brane-localized fields.  Since the Higgs doublets arise as composites of 
the CFT, the AdS interpretation is that they are fields localized to 
the GUT brane.  SU(5) is broken by the strong dynamics of the CFT, 
corresponding to boundary conditions on the GUT brane that break SU(5) 
down to the SM gauge group.  The AdS interpretation of doublet-triplet 
splitting therefore appears qualitatively similar to the flat space model 
of Ref.~\cite{Hebecker:2001wq}.  Namely, Higgs doublets can be present
on the GUT brane because the GUT brane respects just the SM group.

On the CFT side, we find several perturbatively stable vacua, such as
$Q,\Qbar$'s containing no vevs, that translate into SU(5)
preserving boundary conditions of the GUT brane.  However, using the
improved knowledge of strongly-coupled supersymmetric theories, we showed 
that this SU(5) invariant vacuum was non-perturbatively unstable 
(developing a run-away superpotential).  This suggests the perturbatively 
stable SU(5) preserving even boundary conditions on the AdS side 
may actually be unstable non-perturbatively, developing a run-away 
superpotential on the GUT brane.

The remainder of the SU(12)/SU(5) global symmetry of our CFT is explicitly 
broken by Planck scale effects, giving mass for the pseudo-Goldstones 
through $1/\Mpl$-suppressed operators.  These operators are enhanced by
the large anomalous dimension of the CFT to $1/M_{\rm GUT}$.  An
equivalent 4D description corresponds to rewriting the higher dimensional
operators in terms of renormalizable interactions with spectator fields
that get integrated out, as we showed in
Eq.~(\ref{renormalizable-superpotential}).  The AdS interpretation is
that these auxiliary fields are bulk fields with various wavefunction
profiles that lead to both higher dimensional operators involving CFT
fields as well as matter-CFT couplings, i.e., Yukawa couplings.  

Finally, our CFT is a small $N_c$ theory whose breaking effects, 
the massive gauge bosons associated with broken Sp(4) generators, 
correspond in the AdS picture to strongly-coupled Kaluza-Klein resonances 
of the compactified spacetime.  This dual 5D interpretation is, therefore, 
not obviously useful as a calculational tool for us.  It could however 
provide insight into translating strongly-coupled AdS theories into 
small $N$ 4D conformal 
field theories that, for supersymmetric cases at least, we have 
better knowledge.

\section{Discussion}

We have constructed a tripletless product SU(5) $\times$ Sp(4) grand 
unified model by throwing out both the Higgs doublets and triplets
from SU(5).  This resolved the fine-tuning and rapid proton decay problems
associated with Higgs triplets.  Higgs doublets are present in the model,
arising not from SU(5) but instead a four dimensional 
representation of the Sp(4) gauge group.  The Sp(4) group has enough
flavor to be in the conformal window, and therefore its gauge coupling
naturally flows to a strongly-coupled infrared fixed-point at low energy. 
This ensures gauge coupling unification to percent level accuracy, but 
without the coincidence problem of why the Sp(4) group got strong at the 
GUT scale (it was, in fact, always strong).  
Yukawa couplings, including the top, arise through dimension five operators 
that are enhanced by the large anomalous dimension of the Higgs fields.
Proton decay mediated by dimension five operators, 
while absent at the perturbative level,
arises non-perturbatively from the Sp(4) gauge interaction
but is sufficiently suppressed.
With gravity or gaugino mediation, non-universal gaugino masses 
are predicted, satisfying specific one-loop renormalization
group invariant relations.  These predictions can be tested
at the LHC and a linear collider.

It is amusing to note that some recent approaches to mediating
supersymmetry breaking with flavor-blind scalar masses also
employ supersymmetric gauge theories in the conformal window.
Luty and Sundrum proposed a hidden sector with a CFT whose
large anomalous dimensions caused a suppression of the direct couplings 
of the hidden sector to the visible sector \cite{Luty:2001jh},
thereby allowing anomaly mediation to dominate.
Separately, Nelson and Strassler proposed using a CFT to 
generate Yukawa hierarchies as well as suppressing soft-breaking 
parameters \cite{Nelson:2000sn}.  It is tempting to consider
models in which the CFT is used for both product GUT breaking and 
supersymmetry breaking, as this may yield firmer predictions 
for the soft supersymmetry breaking parameters.  However, various
hurdles must be surmounted.  In Luty-Sundrum, for example, a much 
larger hierarchy between the strong and confinement scales is needed
to get enough conformal sequestering, and it certainly appears problematic
to obtain Higgs fields from the CFT (as we do) that would be part of 
the hidden sector (for Luty-Sundrum).

Finally, it is interesting to speculate about other product GUT
theories in the conformal window.  It is straightforward to extend
our model to an SO(10) $\times$ Sp(4), but additional auxiliary
fields transforming under SO(10) are needed to break this group
down the SM\@.  Another possibility is to consider a tripletfull model 
based on SU(5) $\times$ Sp(6). 
This model contains fundamental Higgs doublets and triplets, where
the triplets marry off with a ${\bf 6} = {\bf 3 + \bar{3}}$ 
representation of Sp(6).  More matter is needed so that the Sp(6) 
theory resides in the conformal window, but otherwise the analysis
proceeds analogously to the analysis of the SU(5) $\times$ Sp(4)
model presented above.  Yukawa couplings are simply the ordinary 
MSSM marginal operators.  One important difference, however, 
is that the Higgs doublets can acquire a supersymmetric mass term from 
non-perturbative effects (instantons) of Sp(6).  This is again suppressed 
by a ratio of the dynamical scale to the vev, but now this ratio has to 
be numerically order 10 to ensure this contribution is at or below 
the weak scale.  This may be somewhat tricky to achieve simultaneous with 
a sufficiently strongly-coupled Sp(6) theory that is needed to avoid
large tree-level threshold corrections as well as providing a sufficient
enhancement of the masses of the pseudo-Goldstone fields.
In any case, there is clearly a larger class of product GUT theories 
that live in the conformal window, solving the doublet-triplet splitting 
problem as well as naturally preserving gauge coupling unification.

\section*{Acknowledgments}

We thank B. Feng, O. Lunin, Y. Nomura, E. Poppitz, and T. Watari for
discussions.  This work was supported by funds from the Institute for
Advanced Study and in part by the DOE under contract DE-FG02-90ER40542.


\end{document}